\documentclass[pre,aps,amsmath,longbibliography,amsfonts,amssymb,twocolumn]{revtex4-1}
\usepackage{graphicx,CJK}
\usepackage{xspace}
\renewcommand{\epsilon}{\varepsilon}
\newcommand{\add}{\mu}
\newcommand{\ethr}{\epsilon_{th}}
\newcommand{\narw}{AWL\xspace}
\newcommand{\narwa}{AWLA\xspace}
\newcommand{\narwr}{AWLR\xspace}
\DeclareMathOperator{\sgn}{sgn}
\begin{document}
\begin{CJK*}{UTF8}{mj}
\title{One-dimensional annihilating random walk with long-range interaction}
\author{Su-Chan Park (박수찬)}
\affiliation{Department of Physics, The Catholic University of Korea, Bucheon 14662, Republic of Korea}
\begin{abstract}
We study the annihilating random walk with long-range interaction in one dimension. Each particle performs random walks on a one-dimensional ring in such a way that the probability of hopping toward the nearest particle is $W= [1 - \epsilon (x+\add)^{-\sigma}]/2$ (the probability of moving away from its nearest particle is $1-W$), where $x$ is the distance from the hopping particle to its nearest particle and $\epsilon$, $\add$, and $\sigma$ are parameters. For positive (negative) $\epsilon$, a particle is effectively repulsed (attracted) by its nearest particle and each hopping is generally biased. On encounter, two particles are immediately removed from the system. We first study the survival probability and the mean spreading behaves in the long-time limit if there are only two particles in the beginning. Then, we study how the density decays to zero if all sites are occupied at the outset. We find that the asymptotic behaviors are classified by seven categories: (i) $\sigma>1$ or $\epsilon=0$, (ii) $\sigma = 1$ and $2\epsilon > 1$, (iii) $\sigma=1$ and $2\epsilon = 1$, (iv) $\sigma = 1$ and $2\epsilon < 1$, (v) $\sigma<1$ and $\epsilon > 0$, (vi) $\sigma = 0$ and $\epsilon<0$, and (vii) $0 < \sigma <1$ and $\epsilon<0$. The asymptotic behaviors in each category are universal in the sense that $\add$ (and sometimes $\epsilon$) cannot affect the asymptotic behaviors.
\end{abstract}
\date{\today}
\maketitle
\end{CJK*}
\section{Introduction}
The annihilating random walk  
and its close relative the coalescing random walk describe processes 
whereby diffusing particles react on encounter.
In their presumably simplest setting, particles perform random walks on a $d$-dimensional hypercubic lattice and 
they undergo pairwise annihilation ($A+A\rightarrow 0$) or coalescence ($A+A \rightarrow A$) 
if two particles happen to occupy a same site. 
Due to exact solvability and wide applicability to various fields, these processes have been
studied extensively for many years~\cite{Bramson1980,Toussaint1983,Torney1983,KR1984,Kang1984,KR1985,Lushnikov1986,Peliti1986,Lushnikov1987,Doering1988,Spouge1988,Lee1994,Henkel1995,Henkel1997,Bares1999,PPK2001,PJP2005,bA2005}.

In the generic setting, hopping of each particle is symmetric in the sense that
the direction of hopping is chosen with equal probability among $2d$ nearest-neighbor sites.
In this case, the upper critical dimension $d_c$ is 2 and the asymptotic behavior of particle density 
is universal with $t^{-d/2}$ for $d<d_c$ and $t^{-1}$ for $d>d_c$.

It is quite natural to ask what would happen if hopping is biased. 
In the framework of the field theory~\cite{Doi1976A,Doi1976B,Peliti1985,Peliti1986,Lee1994},
it is easy to understand that global bias does not affect the asymptotic behavior, because
the bias is removed by the Galilean transformation~\cite{PHP2005a}.
By the global bias, we mean that the direction and the strength 
of the bias does not depend on the position of a particle.

If bias varies with position and/or time, then the Galilean transformation cannot remove the bias.
This kind of bias can be relevant in the renormalization group (RG) sense and 
the asymptotic behavior would change.
One way of implementing such a bias is to introduce a quenched noise
in such a way that the strength of the bias varies from site to site~\cite{Schutz1997,Schutz1998,Park1998,Chung1999,Ledoussal1999,Richardson1999,Hnatich2000}.

Recently, another form of bias that cannot be removed by the Galilean transformation was 
introduced~\cite{Sen2015}, initially motivated from opinion dynamics~\cite{Biswas2009,Biswas2011}. 
This hopping bias is implemented
in such a way that a particle prefers hopping toward its nearest particle. 
Unlike the quenched noise, the direction of hopping depends on which configuration the system 
is in and, accordingly, it can change with time. 

In the original setting~\cite{Sen2015}, the strength of the bias 
does not depend on how far  a walker's nearest particle is located. Then the bias is generalized in Ref.~\cite{Park2020B} such that 
the strength of the bias is a decreasing power-law function of the distance from a particle to 
its nearest one. It was found that 
the asymptotic behavior of the density depends on the form of the power-law function.

In this paper, we further generalize the one-dimensional annihilating random walk in Ref.~\cite{Park2020B}
by allowing that a particle is repulsed by its nearest particle.
As we will see, the repulsion triggers rich phenomena. 
In Sec.~\ref{Sec:model}, we define the generalized model and 
introduce two initial conditions that are termed as the two-particle and fully occupied initial
conditions, respectively.
Section~\ref{Sec:Sur} studies the system with the two-particle initial condition, focusing
on survival probability and mean spreading.
Section~\ref{Sec:ARW} studies how the density behaves in the long-time limit if the
system evolves from the fully occupied initial condition.
In Sec.~\ref{Sec:sum}, we summarize the result of the paper.
\section{\label{Sec:model}Model}
The model is defined on a one-dimensional lattice of size $L$ with periodic boundary conditions. 
Each site of this lattice is either occupied by a particle or vacant. 
Multiple occupancy is not allowed. 
We will denote the occupation number at site $i$ by $s_i$, which takes either 1 or 0.
For convenience, we define 
\begin{align}
	\nonumber
	R_i &= \min\{x|s_{i+x}=1, 1\le x \le L\},\\
	L_i &= \min\{x|s_{i-x}=1, 1\le x \le L\},
\end{align}
where $i \pm L$ should be interpreted as $i$ (periodic boundary condition).
In other words, $R_i$ ($L_i$) is the distance from site $i$ to the nearest occupied site
on the right- (left-) hand side.

With transition rate 1, each particle hops to one of its nearest-neighbor sites.
If a particle at site $i$ is to hop, then it must move to either site $i+1$ or site $i-1$.
Probability $W_i$ of hopping to site $i+1$ is (the probability of hopping to site $i-1$ is naturally $1-W_i$)
\begin{align}
	W_i = \frac{1}{2}   +\frac{\epsilon }{2}\sgn(R_i-L_i) (m_i+\add)^{-\sigma},
	\label{Eq:qdef}
\end{align}
where $m_i=\min\{R_i,L_i\}$, $\sgn(x) (\equiv x/|x|)$ is the sign of $x$ with $\sgn(0)=0$, $\sigma \ge 0$,
and $\epsilon$, $\add$ are constants with the 
restriction $ 0\le |\epsilon| < (1+\add)^\sigma$ to ensure $0< W_i < 1$.
A particle is in a sense repulsed (attracted) by its nearest particle 
when $\epsilon$ is positive (negative). 
If a particle happens to jump to a site that is already occupied,
then the two particles are removed in no time (pairwise annihilation).

Since hopping of a particle is significantly influenced (especially for small $\sigma$) 
by its nearest particle even if they are separated by a large distance, 
we refer to the model as the annihilating random walk with long-range interaction (\narw).
As we will see soon, the sign of $\epsilon$ plays a crucial role. To emphasize the effect of the sign, 
we will also refer to the model with positive (negative) $\epsilon$ as the
annihilating random walk with long-range repulsion (attraction), which will be abbreviated
as \narwr (\narwa).

In the following sections, we study the \narw for two initial conditions.
One is the two-particle initial condition in which
there are only two particles in a row at $t=0$ in an infinite system.
In this case, we are interested in the survival probability $S(t)$ that
two particles survive up to time $t$ and
the mean distance $R(t)$ between the two particles, conditioned that they are not annihilated
up to time $t$.
The asymptotic behaviors of $S(t)$ and $R(t)$ will be studied in Sec.~\ref{Sec:Sur}.

The other is the fully occupied initial condition in which $s_i=1$ for all $i$ at $t=0$.
In this case, we investigate the asymptotic behavior of particle density 
\begin{align}
	\rho(t) = \frac{1}{L}\sum_i \langle s_i \rangle,
\end{align}
where $\langle \cdots \rangle$ stands for average over ensemble.
The asymptotic behavior of the density $\rho$ of the \narwa was first reported in Ref.~\cite{Sen2015} 
for $\sigma=0$ and later in Ref.~\cite{Park2020B} for any $\sigma$, which is
\begin{align}
	\rho(t) \sim 
	\begin{cases}
		t^{-1/(1+\sigma)},& \sigma < 1,\\
		t^{-1/2}, & \sigma \ge 1.
	\end{cases}
	\label{Eq:negdel}
\end{align}
Throughout the paper,  we write $f(x) \sim g(x)$ if
\begin{align}
	0< \left | \lim_{x\rightarrow x_0} \frac{f(x)}{g(x)} \right | < \infty,
\end{align}
where $x_0=0$ or $x_0=\infty$, depending on the context.
In Sec.~\ref{Sec:ARW}, we will investigate the asymptotic behavior of $\rho$ for any value of $\epsilon$
and we will reproduce Eq.~\eqref{Eq:negdel} for negative $\epsilon$ in due course.
\section{\label{Sec:Sur}Survival probability and mean spreading}
This section studies how the system evolves if it starts from the two-particle initial condition.
Let $\widetilde P_i(t)$ be the probability that the distance between the two particles is $i$ at time $t$.
$\widetilde P_0(t)$ is the probability that the two particles are annihilated before $t$.
Considering that probability of hopping to the left (right) of the left particle is the same as 
that of hopping to the right (left) of the right particle,
we write the master equation 
\begin{align}
	\frac{1}{2}\frac{\partial \widetilde P_i}{\partial t}
	=  q_{i-1} \widetilde P_{i-1} +  (1-q_{i+1}) \widetilde P_{i+1}-
	 (1-\delta_{i,0} )\widetilde P_i,
\end{align}
where $\delta_{i,j}$ is the Kronecker $\delta$ symbol and
\begin{align}
	q_i = \frac{1}{2} + \frac{\epsilon}{2} (i+\add)^{-\sigma},
	\label{Eq:qidef}
\end{align} 
with $q_0 =q_{-1}=0$.
Defining $P(i,t) = \widetilde P(i, 2t)$, we write
\begin{align}
	\label{Eq:Pimaster}
	\frac{dP_i}{dt}
	&= q_{i-1} P_{i-1} + (1-q_{i+1}) P_{i+1} - \left (1-\delta_{i,0}\right )P_i,
\end{align}
which is equivalent to a random-walk problem with an absorbing wall at the origin, interpreting $i$ to be
a site where the walker is located.
In this section, we study this random walk with the initial condition $P_i(0)=\delta_{i,1}$.

We are interested in the survival probability $S(t)$ and 
the mean spreading $R(t)$ conditioned on survival, defined as
\begin{align}
	S(t) =  1 - P_0(t),\quad
	R(t) = \sum_{n=1}^\infty \frac{n P_n(t)}{S(t)}.
\end{align}
We will denote the probability that the walker never visits the absorbing wall by $P_s$,
which is obtained as
\begin{align}
	P_s = \lim_{t\rightarrow \infty} S(t).
\end{align}

The continuous-time random walk is related to the discrete-time random walk
in the following way.
Let $d_{i,n}$ be the probability that the walker is located at site $i$
after $n$th jump in the discrete-time random walk, which satisfies
\begin{align}
	d_{i,n+1} = q_{i-1} d_{i-1,n} +(1-q_{i+1}) d_{i+1,n}  + \delta_{i,0} d_{0,n},
	\label{Eq:dindef}
\end{align}
with the initial condition $d_i(0) = \delta_{i,1}$.
Since the number of jumps up to time $t$ 
follows the Poisson distribution with mean $t$, $P_i(t)$ can be found by
\begin{align}
	P_i(t) = \sum_{n=0}^\infty \frac{t^n}{n!} e^{-t} d_{i,n},
	\label{Eq:Pin}
\end{align}
which yields
\begin{align}
	\label{Eq:SR}
	S(t) &= \sum_{n=0}^\infty \frac{t^n}{n!} e^{-t}\xi_n ,\quad
	R(t) = \frac{1}{S(t)} \sum_{n=0}^\infty \frac{t^n}{n!} e^{-t} r_n,\\
	\nonumber
	\xi_n &\equiv \sum_{i=1}^\infty d_{i,n},
	\quad\qquad 
	r_n \equiv \sum_{i=1}^\infty i d_{i,n}.
\end{align}

For numerical studies of $S(t)$ and $R(t)$,
we either use Eq.~\eqref{Eq:SR} with numerical calculation of $d_{i,n}$ [especially when
$S(t)$ is extremely small] or perform Monte Carlo simulations 
of the discrete-time random walk [especially when the observation time is large or $S(t)$ at the end of the observation is larger than $10^{-10}$ ].
As long as we are interested in the long-time behavior, whether time is continuous or discrete is 
immaterial in most cases with one exception in this paper.

We begin with investigating the probability $F_i(r)$ that the walker starting from site $i$ visits site $r$ 
at least once. Notice that $P_s$ can be obtained by
\begin{align}
	P_s = \lim_{r\rightarrow \infty} F_1(r).
\end{align}
Due to the Markov property, we have a recursion relation
\begin{align}
	F_i = q_i F_{i+1} + (1-q_i) F_{i-1}.
\end{align}
Since $F_0=0$, we get
\begin{align}
	F_{i+1} - F_i = 
	\left ( F_i - F_{i-1} \right ) 
	\frac{1-q_i}{q_i} = F_1 \prod_{k=1}^i \frac{1-q_k}{q_k},
\end{align}
which, along with $F_r=1$ by definition, gives
\begin{align}
	\label{Eq:F1rform}
	F_n(r) = \frac{G_n}{G_r},\quad
	G_n \equiv  1 + \sum_{i=1}^{n-1} \prod_{k=1}^i \frac{1-\epsilon(k+\add)^{-\sigma}}{1+\epsilon (k+\add)^{-\sigma}},
\end{align}
where $G_0\equiv 0$ and $G_1 \equiv 1$.
For $\epsilon = 0$, we get trivially $G_n = n$ and $F_n(r) = n/r$.

For $\sigma=0$, one can readily get
\begin{align}
	G_n= \frac{1+\epsilon}{2\epsilon}
	\left [ 1 - \left ( \frac{1-\epsilon}{1+\epsilon} \right )^n \right ],
\end{align}
which gives 
\begin{align}
	P_s = \lim_{r\rightarrow\infty} G_r^{-1}=\frac{2\epsilon}{1+\epsilon}\Theta(\epsilon),
	\label{Eq:ps0}
\end{align}
where $\Theta(\cdot)$ is the Heaviside step function.
Note that $G_n$ diverges exponentially with $n$ for $\epsilon < 0$,
which indicates that $F_1(r)$ decreases exponentially with $r$.

For $\sigma=1$, we can write $G_n$ as
\begin{align}
	G_n = \frac{\Gamma(1+\add+\epsilon)}{\Gamma(1+\add-\epsilon)}\sum_{i=0}^{n-1} 
	 \frac{\Gamma(i+1 +\add-\epsilon)}{\Gamma(i+1+\add+\epsilon)},
	 \label{Eq:s1Gn}
\end{align}
where 
$\Gamma(\cdot)$ is the Gamma function.
For $2\epsilon = 1$, we get
\begin{align}
	G_n 
	= \sum_{i=0}^{n-1} \frac{2\add+1}{2i+2\add + 1}
	\sim \frac{2\add+1}{2} \ln n.
	\label{Eq:Gns1half}
\end{align}
For $2\epsilon \neq 1$, we use an identity
\begin{align}
	\label{Eq:diff1}
	(a-b) \frac{\Gamma(x+b)}{\Gamma(x+a+1)}=
	\frac{\Gamma(x+b)}{\Gamma(x+a)}-\frac{\Gamma(x+1+b)}{\Gamma(x+1+a)},
\end{align}
to obtain
\begin{align}
	G_n&= 
	\frac{\Gamma(1+\add+\epsilon)\Gamma(n+1+\add-\epsilon)}{(1-2\epsilon)\Gamma(1+\add-\epsilon)\Gamma(n+\add+\epsilon)} 
	+\frac{\add+\epsilon}{2\epsilon-1}.
	\label{Eq:Gns1}
\end{align}
One can readily find 
\begin{align}
	P_s =\lim_{r\rightarrow\infty}F_1(r)= \frac{2\epsilon-1}{\add + \epsilon} \Theta(2\epsilon -1).
	\label{Eq:Psse}
\end{align}
Unlike the case with $\sigma=0$, $P_s$ can be 0 even if $\epsilon>0$.
For $2\epsilon < 1$, $G_r$ for large $r$ behaves as
\begin{align}
	F_1(r)^{-1} = G_r \sim
		r^{1-2\epsilon}, 
	\label{Eq:Gns1asym}
\end{align}
where we have used the Stirling's formula.
Note that the power in the asymptotic behavior in Eq.~\eqref{Eq:Gns1asym}
varies continuously with $\epsilon$, but does not depend on $\add$.

\begin{figure}
	\includegraphics[width=\linewidth]{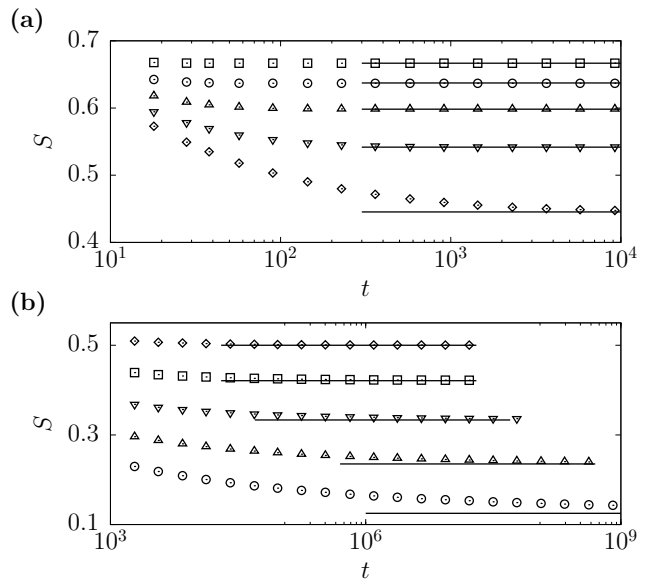}
	\caption{\label{Fig:def} 
	(a) Plots of $S$ vs. $t$
	for $\sigma=0$ (square), 0.2 (circle), 0.4 (up-triangle),
	0.6 (down-triangle), and 0.8 (diamond), top to bottom, on a semilogarithmic scale. Here $\epsilon=0.5$ and $\add=0$.
	Hoizontal line segments show the value of $P_s$ numerically obtained from Eq.~\eqref{Eq:F1rform}.
	(b) Semilogarithmic plots of $S$ vs. $t$ for $\sigma=\add=1$ and
	for $\epsilon=0.6$ (circle), 0.7 (up-triangle),
	0.8 (down-triangle), 0.9 (square), and 1 (diamond), bottom to top.
	Hoizontal line segments indicate the predicted $P_s$ in Eq.~\eqref{Eq:Psse}.
	}
\end{figure}
For $0<\sigma< 1$, we show in Appendix~\ref{App:Gn}
that $G_n$ converges as $n\rightarrow \infty$ as long as $\epsilon>0$.
Hence, we conclude that $P_s$ for any positive $\epsilon$ is nonzero if  $\sigma$ is strictly
smaller than 1. 
Appendix~\ref{App:Gn} also shows that $G_n$ diverges as $n\rightarrow\infty$ 
for any $\epsilon$ if $\sigma>1$, which amounts to $P_s = 0$.
Defining the threshold value as $\ethr \equiv \sup\{\epsilon|P_s = 0\}$,
we obtain
\begin{align}
	\ethr =
	\begin{cases}
		0,& \sigma<1,\\
		1/2, & \sigma=1,\\
		\infty, & \sigma>1.
	\end{cases}
\end{align}

To confirm, we compare Monte Carlo simulation results
with the predictions.
Figure~\ref{Fig:def}(a) compares the simulation results for $\sigma<1$ 
to the corresponding prediction, to show perfect agreement.
In Fig.~\ref{Fig:def}(b), we present the simulation results 
for $\sigma=1$ and $2\epsilon >1$
to find that $P_s$ in Eq.~\eqref{Eq:Psse}
is in perfect agreement with simulations in the long-time limit.

We now present an approximate expression for $G_n$.
Since we are mainly interested in how $G_n$ behaves for large $n$, 
we expect that the main contribution of the sum in Eq.~\eqref{Eq:F1rform} occurs when $i$ is 
large.
Accordingly, we have an approximation
\begin{align}
	G_n \approx \int_1^n dx 
	e^{ -2 \epsilon I(x;\sigma)},\quad
	I(x;\sigma) \equiv \int_{1+\add}^{x+\add} y^{-\sigma} dy,
	\label{Eq:GnIx}
\end{align}
where we used $\ln [(1-x)/(1+x) ] \simeq -2 x$ 
and replaced sums with integrals.

For $\sigma>1$, $I(x;\sigma)$ converges as $x\rightarrow\infty$,
which yields $G_n \sim n$ for any $\epsilon$, as also shown in Appendix~\ref{App:Gn}.
Since $G_n=n$ for $\epsilon=0$ (unbiased case), we conclude
that the case with $\sigma>1$ shares the universal asymptotic behavior with the unbiased
random walk. We will arrive at the same conclusion when we discuss the asymptotic behavior of $R(t)$
and $S(t)$.

Since $I(x;\sigma)\sim x^{1-\sigma}$ for $\sigma<1$,
$G_n$ for $\epsilon>0$ is bounded as expected.
For $\epsilon<0$, we obtain the asymptotic behavior of $F_1(r)$ as
\begin{align}
	F_1(r) &= G_r^{-1} \sim r^{-\sigma} \exp \left ( -\frac{2|\epsilon|}{1-\sigma}
	r^{1-\sigma} \right ),
	\label{Eq:Gnslt1}
\end{align}
where we have used Eq.~\eqref{Eq:asym} in Appendix~\ref{App:asym}.
For $\sigma=1$, one can easily check that Eq.~\eqref{Eq:GnIx} gives the same asymptotic behaviors
as Eqs.~\eqref{Eq:Gns1half} and \eqref{Eq:Gns1asym}.

Now we will find the asymptotic behaviors of $R(t)$ and $S(t)$.
Our analysis of $R(t)$ begins with writing down an equation for $R(t)$. 
Using the master equation~\eqref{Eq:Pimaster}, we get
\begin{align}
	\frac{dR}{dt} = 
	 \epsilon \sum_{n=1}^\infty (n+\add)^{-\sigma} \psi_n(t) - R(t)\frac{d\ln S(t)}{dt},
\end{align}
where $\psi_n(t) \equiv P_n(t) /S(t)$ with $\sum_{n=1}^\infty \psi_n(t) = 1$.
If we define $u(t) = R(t) S(t)$, then we get 
\begin{align}
	\frac{du}{dt} = 
	 \epsilon S(t) \sum_{n=1}^\infty (n+\add)^{-\sigma} \psi_n(t).
	 \label{Eq:Ut}
\end{align}
Actually, $u(t)$ is the mean distance from the wall to the walker 
that is averaged over \emph{all} ensemble at time $t$.

We find a formal solution for $\sigma=0$ as
\begin{align}
	R(t;\sigma=0) = \frac{1}{S(t)}\left [ R_0 + \epsilon \int_0^t S(t') dt'\right ] ,
\end{align}
where $R_0$ is a constant determined by the 
initial condition ($R_0=1$ for the two-particle initial condition).
Since $S(t)$ saturate to nonzero $P_s$ for positive $\epsilon$ , we find
\begin{align}
	R(t;\sigma=0) \sim \epsilon t.
	\label{Eq:Rs0asym}
\end{align}
Since $S(t) \sim t^{-1/2}$ for $\epsilon = 0$ (see, for example, Ref.~\cite{FellerI}), 
we get $R(t;\epsilon=0) \sim t^{1/2}$.

As we have shown above, $S(t)$ converges to nonzero $P_s$ if 
$\sigma<1$ with positive $\epsilon$ or if $\sigma=1$ with $2\epsilon>1$.
In these cases, we can neglect the second term in the long-time
limit and we have
\begin{align}
\frac{dR}{dt} \approx 
	 \epsilon \sum_n (n+\add)^{-\sigma} \psi_n(t),
	 \label{Eq:Rtact}
\end{align}
which suggests that $R(t)$, not surprisingly, should increase indefinitely.
To find the asymptotic behavior of $R(t)$ for nonzero $P_s$, let us
assume that $\psi_n(t)$ is sharply peaked around $n = R(t)$.
Under this assumption, we can approximate the summation in Eq.~\eqref{Eq:Rtact}
as (we neglect $\add$ because $R$ is large)
\begin{align}
	\sum_n \frac{\psi_n(t)}{(R+\Delta n)^\sigma}
	\approx \frac{1}{R^\sigma} \left [ 1 
	+ \frac{\sigma(\sigma+1)}{2R^2}\left \langle  (\Delta n)^2 \right \rangle_s \right ],
\end{align}
where $R = \langle n \rangle_s$, $\Delta n \equiv n -R$, and $\langle \cdots \rangle_s$ stands for
the average over $\psi_n$.
Hence, we have an approximate equation for $R(t)$ as
\begin{align}
	\frac{dR}{dt}
	\approx \frac{\epsilon}{R^{\sigma}} \left [ 1 + \frac{\sigma(\sigma+1)}{2}\frac{\langle (\Delta n)^2\rangle_s}{R^2} \right ].
	\label{Eq:Rapp}
\end{align}
Neglecting the fluctuation $(\Delta n)^2$, we obtain
\begin{align}
	\frac{dR}{dt}  \approx \epsilon R^{-\sigma} \rightarrow R \approx \left [ \epsilon (1+\sigma) t
	\right ]^{1/(1+\sigma)},
	\label{Eq:Rt}
\end{align}
which reproduces the exact asymptotic behavior \eqref{Eq:Rs0asym} for $\sigma=0$.

Now we argue that keeping only the leading term gives the exact asymptotic behavior
for  $\sigma <1$ and $\epsilon>0$.
A (naive) continuum limit for the master equation yields the Fokker-Planck equation
\begin{align}
	\frac{dP(x;t)}{dt} = - \frac{\partial}{\partial x} \left [ 
	\frac{\epsilon}{x^\sigma} P(x;t) \right ]
	+ \frac{1}{2} \frac{\partial^2}{\partial x^2} P(x;t),
	\label{Eq:FPE}
\end{align}
where $x$ is the continuum version of the site index and we neglect $\add$, assuming $x$ is large.
Since the diffusion term in Eq.~\eqref{Eq:FPE} does not depend on $\epsilon$, 
we expect that the variance of $x$ increases linearly just like the unbiased random
walks.  Accordingly, $\langle \Delta n^2
\rangle_s / R^2 \rightarrow 0$ as $t\rightarrow 0$ for $\sigma<1$ and, in turn, the approximation 
\eqref{Eq:Rt} becomes accurate in the long-time limit; see 
Ref.~\cite{Park2020B} for a similar discussion with negative $\epsilon$.

\begin{figure}
	\includegraphics[width=\linewidth]{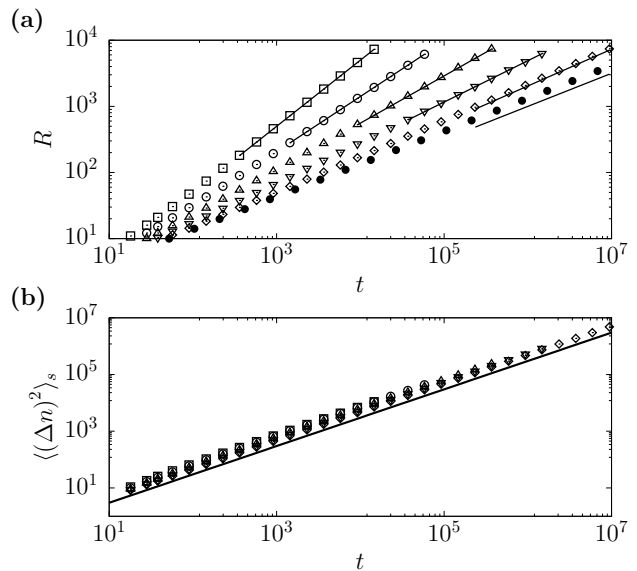}
	\caption{\label{Fig:def2} 
	(a) Double logarithmic plots of $R$ vs. $t$ for $\sigma=0$ (square), 0.2 (circle), 0.4 (up-triangle),
	0.6 (down-triangle), 0.8 (diamond), and 1 (filled circle), top to bottom. 
	Here $\epsilon= 0.5$ for all cases, but $\add=0$ for $\sigma<1$ as in Fig.~\ref{Fig:def}(a) and $\add=1$ for $\sigma=1$ as in Fig.~\ref{Fig:def}(b).
	Line segments depicts the predicted asymptotic behavior~\eqref{Eq:Rt}, which shows
	perfect agreement for $\sigma<1$. The deviation for $\sigma=1$ is discussed in the text.
	(b) Double logarithmic plots of $\langle (\Delta n)^2 \rangle_s$ vs. $t$ for 
	$\sigma=0$ (square), 0.2 (circle), 0.4 (up-triangle), 0.6 (down-triangle), and 0.8 (diamond). 
	A straight line with slope 1 is drawn for guides to the eyes.
	}
\end{figure}

We compare Eq.~\eqref{Eq:Rt} with numerical simulations in Fig.~\ref{Fig:def2}(a). 
Our prediction is in full accord with the simulation results for $\sigma<1$ and $\epsilon>0$.
We also measured the fluctuations in simulations to find that it indeed behaves as 
$\langle (\Delta n)^2 \rangle_s \sim t$ for $\sigma < 1$; see Fig.~\ref{Fig:def2}(b).

In Fig.~\ref{Fig:def2}(a), we also present simulation results for $\sigma=1$ and $2 \epsilon=1$ with
comparison to Eq.~\eqref{Eq:Rt}.
Although the power is still consistent with the prediction,
the coefficient deviates from the prediction.
Since $R^2 \sim \langle (\Delta n)^2 \rangle_s \sim t$ for $\sigma=1$ (and $2\epsilon>1$),
we cannot simply neglect the fluctuation $(\Delta n)^2$, but
it only increases the coefficient, which explains why
Eq.~\eqref{Eq:Rt} lies below the simulation data for $\sigma=1$ in Fig.~\ref{Fig:def2}(a).

Let us continue investigating the case with $\sigma=1$ for arbitrary $\epsilon$.
As above, we begin with writing down an approximate equation for $R(t)$ as
\begin{align}
	\frac{dR}{dt} 
	\approx
	 \frac{\epsilon}{R} - \frac{d \ln S(t)}{d t} R,
\end{align}
where we again neglected the fluctuation. For later purposes, we also write down an approximate equation for $u$,
\begin{align}
	\frac{d u^2}{dt} \approx 2\epsilon S(t)^2.
	\label{Eq:uS}
\end{align}
Notice that Eq.~\eqref{Eq:uS} again predicts  $R\sim u \sim \sqrt{t}$
for $2\epsilon >1$, because $S(t)$ in this case saturates to a nonzero value.

Until now, we have investigated the cases with $P_s>0$.
To find $S$ and $R$ for $P_s=0$, we will use the following relation.
If $S(t) \rightarrow 0$ while $R(t) \rightarrow \infty$, 
then $S$ and $R$ are related by
\begin{align}
S(t) \approx F_1(R),
	\label{Eq:Stnege}
\end{align}
because surviving samples typically arrive at $R(t)$ at time $t$. 
We will repeatedly use Eq.~\eqref{Eq:Stnege} in what follows.

We will find the asymptotic behaviors of $S$ and $R$ for $2\epsilon \le 1$ in a self-consistent manner. 
We first assume $0 \le 2\epsilon < 1$.
Since the repulsion gets stronger as $\epsilon$ gets larger, 
it seems plausible to expect that $R(t)$ should be a nondecreasing function of $\epsilon$
for given $t$ and, in turn, $R(t) \sim \sqrt{t}$ for $\epsilon \ge 0$,
because $R(t) \sim \sqrt{t}$ not only for $\epsilon = 0$ but also for $2\epsilon >1$.

Using Eqs~\eqref{Eq:Stnege} and \eqref{Eq:Gns1asym} for $0 < 2\epsilon < 1$, we find 
\begin{align}
S(t) \sim t^{-(1-2\epsilon)/2}. 
	\label{Eq:Stneges1}
\end{align}
If we plug Eq.~\eqref{Eq:Stneges1} into Eq.~\eqref{Eq:uS},
then we get $u\sim t^\epsilon$, which consistently gives
$R = u/S \sim \sqrt{t}$.
Note that $u(t)$ diverges for $0< 2\epsilon < 1$
even though $S(t) \rightarrow 0$ as $t\rightarrow \infty$.

Since $F_1(r) \sim 1/\ln r$ for $2\epsilon = 1$, Eq.~\eqref{Eq:Stnege} along
with Eq.~\eqref{Eq:Gns1half} gives 
\begin{align}
	S(t) \sim 1/\ln t.
\end{align}
Therefore, we get
\begin{align}
	u^2 \sim \int^t \frac{dt}{(\ln t)^2}
	= \int^x \frac{e^x}{x^2} dx
	\sim \frac{e^x}{x^2} = \frac{t}{(\ln t)^2},
\end{align}
where we made a change of variables
$x = \ln t$ and Eq.~\eqref{Eq:asym} was used.
The logarithm correction in $u$ neatly disappears in the leading behavior
of $R(t)$ and we get $R(t) \sim \sqrt{t}$ for all positive $\epsilon$.
This is consistent with the numerical observation in Fig.~\ref{Fig:def2}(a)
and the assumption that $R(t)$ is a nondecreasing function of $\epsilon$.

For negative $\epsilon$, Eq.~\eqref{Eq:Ut} shows that 
$u(t)$ always decreases regardless of the initial condition, which shows
$u(t) \rightarrow 0$ as $t\rightarrow \infty$.
Assuming that Eq.~\eqref{Eq:uS} is a valid approximation
for negative $\epsilon$, we get
\begin{align}
	u^2(t) = \int_\infty^t \frac{d u^2}{dt'} dt'
	\approx  2 |\epsilon| \int_t^\infty S(t')^2 dt'.
	\label{Eq:u2S}
\end{align}
Assuming $R(t) \sim t^\gamma$ for negative $\epsilon$
and using Eq.~\eqref{Eq:Stnege},
we get
\begin{align}
	S(t) \sim t^{-(1-2\epsilon)\gamma},
\end{align}
which together with Eq.~\eqref{Eq:u2S} gives
\begin{align}
	u(t) \sim t^{-(1-2\epsilon)\gamma+1/2}.
\end{align}
Since $R(t) = u(t)/ S(t)$, we get the self-consistent
solution $\gamma =1/2$, that is, $R\sim \sqrt{t}$.

Our findings for $\sigma=1$ are summarized as
\begin{align}
	R(t) \sim \sqrt{t},\quad
	S(t) \sim \begin{cases}
		(2\epsilon-1)/(\add+\epsilon),& 2\epsilon>1,\\
		1/\ln t, & 2\epsilon =1,\\
		t^{-(1-2\epsilon)/2}, & 2\epsilon<1.
	\end{cases}
	\label{Eq:RSs1}
\end{align}
Since $R(t) \sim \sqrt{t}$, neglect of the fluctuation
only affects the coefficient and the approximate equation 
is expected to give the correct power-law behavior.

\begin{figure}
	\includegraphics[width=\linewidth]{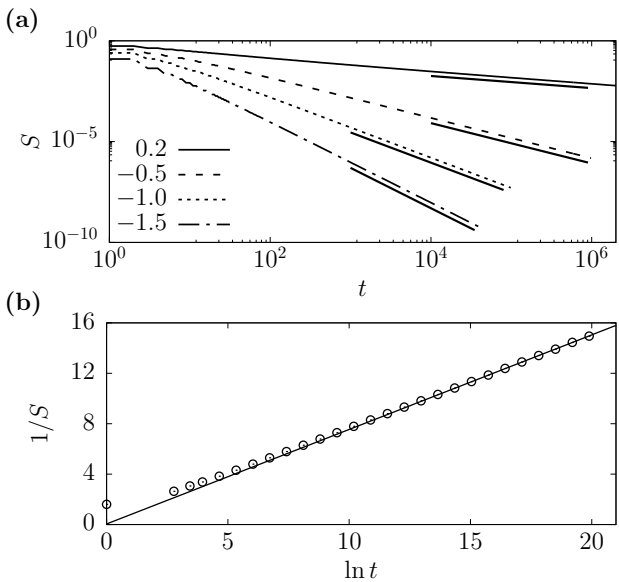}
	\caption{\label{Fig:def1} 
	Survival probability for $\sigma=1$ and $\mu=1$.
	(a) Double logarithmic plots of $S$ vs. $t$
	for $\epsilon=0.2$, $-0.5$, $-1$, and $-1.5$, top to bottom. 
	Line segments with slope $\epsilon - \frac{1}{2}$ are 
	for guides to the eyes. 
	(b) Plot of $1/S$ vs. $\ln t$ for $\epsilon = 0.5$.
	The straight line depicts a function $\frac{3}{4} \ln t + a$,
	where $a$ is determined by a fitting in the region $\ln t \ge 16$;
	see \eqref{Eq:Ss1pred}.
	}
\end{figure}

To support the prediction~\eqref{Eq:RSs1} for $\sigma=1$, we performed
Monte Carlo simulations. In the simulations, we set $\add=1$. In Fig.~\ref{Fig:def1}(a), we 
depict $S(t)$ for $\epsilon = 0.2$, $-0.5$, $-1$, and $-1.5$
on a double logarithmic scale, together with the predicted asymptotic behavior
\eqref{Eq:RSs1} for $2\epsilon<1$ as line segments. The prediction 
perfectly explains the data.
For $2\epsilon =1$, we put $n\sim \sqrt{t}$ in Eq.~\eqref{Eq:Gns1half} to get
\begin{align}
	S(t)^{-1} \sim \frac{3}{4} \ln t.
	\label{Eq:Ss1pred}
\end{align}
In Fig.~\ref{Fig:def1}(b), 
simulation results are compared to the prediction~\eqref{Eq:Ss1pred}
to show excellent agreement.

It is worth while to mention that 
De Coninck \textit{et al.}~\cite{Deconinck2008} studied a similar random walk 
with a reflecting wall at the origin. 
The hopping probability in Ref.~\cite{Deconinck2008} is the same as ours if we set $\sigma=1$ 
and $\add=-\epsilon = \delta/2$.
When the wall at the origin is reflecting, De Coninck \textit{et al.}~\cite{Deconinck2008} found  
$R(t) \sim t^{1-\delta/2}$ for $1 < \delta < 2$, which varies continuously with $\delta$.
Since $\add = -\epsilon$ in Ref.~\cite{Deconinck2008}, it is unclear
whether the exponent depends on $\epsilon$ or $\add$ or both.
Our results seem to suggest that only $\epsilon$ governs the universal behavior,
but detailed analyses are requested for further understanding of the random walk
with the reflecting wall, which is beyond the scope of the present paper.

Since $R(t) \sim \sqrt{t}$ if $\sigma=1$ or if $\sigma=\infty$,
it is natural to expect that $R(t) \sim \sqrt{t}$ for any $\sigma> 1$.
In this case, the term with $R^{-\sigma}$
in Eq.~\eqref{Eq:Rt} is negligible and 
we get $S(t) \sim R(t) \sim t^{-1/2}$.
Notice that this is also consistent with Eq.~\eqref{Eq:Stnege} because $F_1(r) \sim 1/r$
for $\sigma>1$. Hence, the bias is immaterial if $\sigma>1$
and the long-range nature is crucial only when $\sigma\le 1$.

Last, we investigate the case with $\sigma<1$ and $\epsilon=-|\epsilon|$.
In Appendix~\ref{App:s0att}, we find the exact expression of $P_i(t)$ 
for $\sigma=0$, which is
\begin{align}
	\label{Eq:Pits0}
	P_i(t) = w^{i-1}\frac{2 i}{x} 
	  I_i(x)e^{-t},\,
	w \equiv \sqrt{\frac{1+\epsilon}{1-\epsilon}},
\end{align}
where $x = t\sqrt{1-\epsilon^2}$ and $I_i(x)$ is the modified Bessel function of the first kind.
Using $I_i(x) \sim e^x/\sqrt{2\pi x}$ for large $x$, 
one can readily get
\begin{align}
	S(t) &\sim t^{-3/2} \exp \left [ -\left ( 1 - \sqrt{1-\epsilon^2} \right ) t\right ],\\
	\nonumber
	\pi_i &\equiv \lim_{t\rightarrow \infty} \psi_i(t)= (1-w)^2 i w^{i-1},\\
	\nonumber
	\lim_{t\rightarrow \infty} R(t)&= \frac{1+w}{1-w} = \frac{1+\sqrt{1-\epsilon^2}}{|\epsilon|}.
\end{align}
Note that $\pi_i$ is the quasistationary distribution in that 
it is the steady-state solution of the equation
\begin{align}
	\frac{d \psi_i}{dt} =
	q \psi_{i-1} + ( 1 - q ) \psi_{i+1} +[(1-q)\psi_1 -1 ] 
	\psi_i,
\end{align}
where $\psi_0=0$ and $w^2 = q/(1-q)$. 
For the discrete time random walk, there is no quasistationary state in that
\begin{align}
	\lim_{m\rightarrow \infty} \frac{r_{2m}}{r_{2m-1}} \neq 1,
\end{align}
though $\xi_{2m-1} = \xi_{2m}$ for all $m\ge 1$.

Since $F_1(r)$ decays exponentially for $\sigma<1$, it is plausible to anticipate that
$S(t)$ also decays exponentially in the form
$S(t) \sim t^{-\alpha} \exp(-\lambda t^\beta)$.
If we further assume  $\langle (n+\add)^{-\sigma} \rangle_s \sim t^{-\eta}$, 
then  Eq.~\eqref{Eq:Ut} gives
\begin{align}
	u(t)&= |\epsilon| \int_t^\infty S(t') \left \langle (n+\add)^{-\sigma} \right \rangle_s dt'\\
	& \sim \int_t^\infty x^{-\alpha-\eta} \exp \left ( -\lambda x^\beta\right) dx \nonumber
	\\
	&\sim \int_{t^\beta}^\infty 
	y^{-1+(1-\alpha-\eta)/\beta} e^{-\lambda y} dy
	\sim t^{1-\eta-\beta - \alpha}e^{-\lambda t^\beta}\nonumber
\end{align}
and $ R = u/S \sim t^{1-\eta - \beta}$. 
If a quasistationary state exists, then $\eta$ must be zero by definition
and, in turn, $\beta$ must be 1.
Hence, a quasistationary state cannot exist if $\beta<1$.

\begin{figure}
	\includegraphics[width=\linewidth]{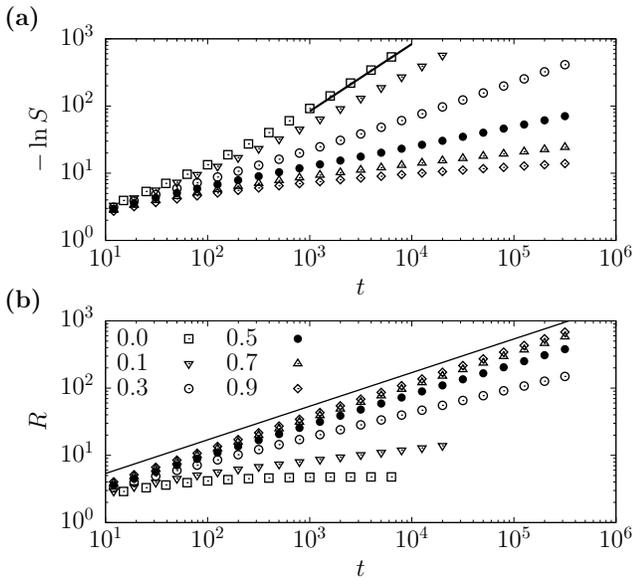}
	\caption{\label{Fig:surex} 
	(a) Double logarithmic plots of $-\ln S$ vs. $t$
	for $\sigma=0$, $0.1$, $0.3$, $0.5$, $0.7$, and $0.9$ (top to bottom)
	with $\epsilon = -0.4$ and $\add=0$. 	
	The line segment depcits the exact asymptotic behavior, $(1-\sqrt{1-\epsilon^2})t$, for $\sigma=0$.
	(b) Double logarithmic plots of $R$ vs. $t$ 
	for $\sigma=0$, $0.1$, $0.3$, $0.5$, $0.7$, and $0.9$ (bottom to top)
	with $\epsilon = -0.4$ and $\add=0$.
	The straight line with slope $0.5$ is a guide for the eyes.
	}
\end{figure}

In Fig.~\ref{Fig:surex}, we present numerical calculations of $S(t)$ and $R(t)$.
As can be seen in Fig.~\ref{Fig:surex}(a), $\beta$ is clearly smaller than 1 for $\sigma>0$
[a fitting of the data for $\sigma = 0.1$ in Fig.~\ref{Fig:surex}(a) gives $\beta\approx 0.8$]
and indeed $R(t)$ increases algebraically; see Fig.~\ref{Fig:surex}(b).
The quasistationary state is a special feature of the case with $\sigma=0$.

Since $R$ increases indefinitely for $0<\sigma<1$,
we use the same logic as in Eq.~\eqref{Eq:Stnege} to get the self-consistent solution
\begin{align}
	(1-\eta-\beta)(1-\sigma) = \beta
	\rightarrow 
	\beta = \frac{1-\sigma}{2-\sigma} (1-\eta),\\
	R \sim t^{(1-\eta)/(2-\sigma)},\quad \alpha = \frac{\sigma}{2-\sigma}(1-\eta).
\end{align}

If we can approximate $\langle (n+\add)^{-\sigma} \rangle_s \propto R^{-\sigma}$ as before,
then the self-consistent argument gives
\begin{align}
	\eta = \frac{\sigma}{2},\quad R\sim \sqrt{t},\quad \beta= \frac{1-\sigma}{2},
	\label{Eq:BDmf}
\end{align}
which cannot be consistent with Fig.~\ref{Fig:surex}(a) especially for small $\sigma$. 
Hence, the mean-field-like approximation 
$\langle (n+\add)^{-\sigma} \rangle_s \propto R^{-\sigma}$ does not work in this case.
Still, Eq.~\eqref{Eq:BDmf} gives a reasonably good approximation for large $\sigma$.
It seems challenging to find the correct asymptotic behavior for $0<\sigma<1$ and $\epsilon<0$, 
which is deferred to a later publication.

\section{\label{Sec:ARW}behavior of the density}
With the two-particle initial condition, the direction of the bias does not change and the particles 
can survive forever with nonzero probability $P_s$, once the repulsion is strong.
When the density is finite, however, 
a particle should meet another particle and is annihilated almost surely 
even if $P_s$ is nonzero and the system size is infinite.
Hence the asymptotic behavior of the density cannot be directly explained by 
the results in Sec.~\ref{Sec:Sur}.
The purpose of this section is to investigate how the density $\rho(t)$ decreases if the system 
evolves from the fully occupied initial condition.

Assume that there are $N=\rho L$ particles at time $t$. Here, $\rho$ is assumed small.
The site index of the $k$th particle is denoted by $n_k$ ($k=1,\ldots,N$, $n_1< n_2<\ldots <n_N$).
The mean distance between the $k$th and $(k+1)$st particles 
is $\langle n_{k+1} - n_k \rangle = 1/\rho$ and the variance 
is expected to be $\langle (n_{k+1} - n_k - 1/\rho )^2 \rangle \sim 1/\rho$ 
(as we will see soon, the exact form of the variance is immaterial as long as its square root is much 
smaller than $1/\rho$).

When $\rho$ is small, a (mean) time gap between any two consecutive pair-annihilation events 
within a region of size $O(1/\rho)$ is 
expected to be large. Assume that the $k$th particle is to be annihilated. As an approximation, we assume that
only the $k$th particle performs random walks and all other particles remain still before the $k$th particle 
is annihilated. Under this approximation, dynamics of the $k$th particle can be mapped
to a random-walk problem with two walls,  one of which is reflecting and the other is absorbing.

To be concrete, let $n_k=\ell$, $n_{k-1}=-r$, and $n_{k+1}=r$, 
where $|r-1/\rho| = O(1/\sqrt{\rho})$ and $\ell =  O(1/\sqrt{\rho})$.
Within the approximation, $n_{k+1}$ and $n_{k-1}$ do not change and
$n_k$ changes according to the rule in Eq.~\eqref{Eq:qdef}.
Since the dynamics are invariant under the transformation $\ell \mapsto -\ell$
and $n_{k+1} \leftrightarrow n_{k-1}$, 
we can set $\ell \ge 0$ without loss of generality and
we can treat the origin as a reflecting wall and site $r$ as an (immovable) absorbing wall. 
In the following, we will call the $k$th particle the walker. 

This random-walk problem can be formulated as follows.
Let $H_i(t)$ be the probability that the walker is located at site $i$ at time $t$
and $H_i(0) = \delta_{i,\ell}$. The time $t$ here should not be confused with the time 
that appeared in the beginning of this section.
We write the master equation ($0 \le i \le r$)
\begin{align}
	\frac{d H_i(t)}{dt}
	= b_{i-1} H_{i-1} + d_{i+1} H_{i+1} - (1-\delta_{r,i})H_i,
\end{align}
where $b_0=1$, $b_{-1}=d_0=d_{r+1}=0$, and ($0 < k < r$)
\begin{align}
	b_k = \frac{1}{2}- \frac{\epsilon}{2} (r-k+\add)^{-\sigma},
	\quad d_k = 1 - b_k.
	\label{Eq:bd}
\end{align}
Recall that the absorbing wall is a particle in the \narw; it
exerts repulsive (attractive) interaction to the walker if $\epsilon$ is positive (negative).
We are interested in the mean first-passage time,
to be denoted by $\tau(\rho)$, for the walker to reach the absorbing wall.

Since $2/\tau(\rho)$ can be interpreted as a rate of removal per particle in the \narw (the factor 2 
is multiplied because of the pair annihilation, but this factor does not affect 
the universal behavior that we will find),
the behavior of $\rho(t)$ can be analyzed by the equation
\begin{align}
	\frac{d\rho}{dt} \propto - \frac{\rho}{\tau(\rho)}.
	\label{Eq:drdt}
\end{align}
If we find $\tau(\rho)$, then we can obtain the asymptotic behavior of $\rho(t)$.
One can use Eq.~\eqref{Eq:drdt} even if the particles perform coalescing random walks ($A+A\rightarrow
A$).

Let $T_i$ be the mean first-passage time if the walker starts from site $i$ at $t=0$.
By definition, we have $T_r = 0$. We will approximate $\tau(\rho)$ as $T_\ell$ 
with $\ell=O(1/\sqrt{\rho})$. Due to the Markov property, we have the recursion relation
\begin{align}
	T_i = 1 + b_i T_{i+1} + d_i T_{i-1}.
	\label{Eq:MFP}
\end{align}
In other words, the walker waits unit time on average and then jumps to site $i+1$ ($i-1$) 
with probability $b_i$ ($d_i$), after which it should spend $T_{i+1}$ ($T_{i-1}$).

To find a formal solution, 
we define $\chi_i \equiv T_i - T_{i+1}$ and we rewrite Eq.~\eqref{Eq:MFP} as
\begin{align}
	\chi_i = \frac{d_i}{b_i} \chi_{i-1} + \frac{1}{b_i}.
	\label{Eq:Si}
\end{align}
Multiplying 
Eq.~\eqref{Eq:Si} by $\prod_{k=1}^{i} (b_k/d_k)$, we get
\begin{align}
	\chi_i \prod_{k=1}^i \frac{b_k}{d_k} -
	\chi_{i-1} \prod_{k=1}^{i-1} \frac{b_k}{d_k} 
	= \frac{1}{d_i} \prod_{k=1}^{i-1} \frac{b_k}{d_k},
	\label{Eq:chii}
\end{align}
where we assume $\prod_{k=1}^0 \equiv  1$.
After a little algebra, we have
\begin{align}
	\chi_{r-n} = \prod_{k=n}^{r-1} \frac{d_{r-k}}{b_{r-k}} + \sum_{j=n}^{r-1} 
	\frac{1}{d_{r-j}} \prod_{k=n}^{j} \frac{d_{r-k}}{b_{r-k}}.
\end{align}
Since $T_{r}=0$, we can write
\begin{align}
	T_i = \sum_{n=i}^{r-1} \chi_n = \sum_{n=1}^{r-i} \chi_{r-n},
	\label{Eq:Tchi}
\end{align}
which gives
\begin{align}
	\label{Eq:Tisol}
	T_i 
	&= \sum_{n=1}^{r-i} 
	\sum_{j=n}^{r-1} \left ( \delta_{j,r-1}+\frac{1}{d_{r-j}} \right ) \prod_{k=n}^j \frac{d_{r-k}}{b_{r-k}}
\end{align}
for $i\ge 1$ and $T_0 = 1+T_1$.

For certain cases, we find a simple expression of $T_0$.
For $\epsilon = 0$ (or equivalently $\sigma=\infty$ with $\add>0$) one can readily get $T_0 = r^2$. 
For later purposes, we write 
\begin{align}
T_0(\sigma=\infty) \sim r^2.
\end{align}
For $\sigma=0$, it is straightforward to get
\begin{align}
	T_0(\sigma=0) = \frac{1-\epsilon^2}{2\epsilon^2} \left [ \left ( \frac{1+\epsilon}{1-\epsilon}
	\right )^r - 1 \right ] 
	 -\frac{r}{\epsilon}.
\end{align}
If $\epsilon$ is positive, then $T_0$ grows exponentially with $r$.
If $\epsilon$ is negative, then $T_0 \sim r$ for large $r$.

By definition, $T_i$ cannot be smaller than $r-i$, so $T_i$ for any case increases indefinitely with $r$
as long as $i \ll r$. 
Considering $0<1-|\epsilon|(1+\add)^{-\sigma} < 2 d_{j} < 2$ for all positive $j$,
we can write 
\begin{align}
	\label{Eq:Ttau}
	T_i 
	\sim \sum_{n=1}^{r-i} 
	\sum_{j=n}^{r-1} \prod_{k=n}^j \frac{1+\epsilon(k+\add)^{-\sigma}}{1-\epsilon(k+\add)^{-\sigma}}.
\end{align}
Since the leading asymptotic behavior of $T_i$ for
large $r$ does not depend on $i$ if $i/r \rightarrow 0$,
it is sufficient to analyze the asymptotic behavior of $T_0$,
\begin{align}
	T_0\sim
	 \sum_{i=2}^{r-1}  \sum_{n=1}^{i} \prod_{k=n}^{i} 
	\frac{ 1+\epsilon(k+\add)^{-\sigma}}{1-\epsilon (k+\add)^{-\sigma}},
	\label{Eq:T0ana}
\end{align}
where we replaced the dummy index $j$ with $i$ and we changed the order of the summations.
For convenience, we neglect the contribution from $i=1$, which does not have $r$ dependence.
In the following three subsections, we will study the \narw for three different cases: $\sigma<1$, $\sigma=1$, and $\sigma>1$.

\subsection{\label{Sec:sigmalt1}$0< \sigma < 1$}
Since $T_0$ diverges with $r$, the dominant contribution to $T_0$
should arise for large $k$ in Eq.~\eqref{Eq:T0ana}. Accordingly, we approximate the product in Eq.~\eqref{Eq:T0ana} as
\begin{align}
	\prod_{k=n}^{i} \frac{1 + \epsilon(k+\add)^{-\sigma}}{1 - \epsilon(k+\add)^{-\sigma}}
	\approx \exp \left ( \int_{n+\add}^{i+\add} 2 \epsilon k^{-\sigma} dk \right ).
	\label{Eq:prod_app}
\end{align} 
Approximating the summations in Eq.~\eqref{Eq:T0ana} by integrals as well, 
we get
\begin{align}
	T_0 &\sim \int_2^{r-1}dx e^{f(x+\add)}
	\int_1^x dn e^{-f(n+\add)},
	\label{Eq:T0lt1Int}
\end{align}
where $f(x) = C_\sigma x^{1-\sigma}$
with $C_\sigma=2\epsilon/(1-\sigma)$.
Since, for $\epsilon>0$ ($C_\sigma >0$) and $x\ge 2$,
\begin{align}
	\nonumber
	\int_1^2 &dn e^{-f(n+\add)} \le \int_1^x dne^{-f(n+\add)} \\
	&\le  \int_0^\infty dne^{-f(n)} =C_\sigma^{1/(\sigma-1)} \Gamma\left (\frac{2-\sigma}{1-\sigma}\right),
\end{align}
we get
\begin{align}
	T_0 \sim \int_1^{r-1} \exp \left ( C_\sigma x^{1-\sigma}\right )dx
	\sim r^{\sigma} \exp \left (C_\sigma r^{1-\sigma}\right ),
\end{align}
where we have used Eq.~\eqref{Eq:asym}.

For negative $\epsilon$ ($C_\sigma<0$), the integral with variable $n$ in Eq.~\eqref{Eq:T0lt1Int}
diverges as $x\rightarrow \infty$.
We again use Eq.~\eqref{Eq:asym} to get
\begin{align}
	T_0 \sim \int_1^{r-1} x^\sigma dx \sim r^{1+\sigma}.
\end{align}

To summarize, we obtain
\begin{align}
	T_0 \sim \begin{cases}
		r^{\sigma} \exp \left (C_\sigma r^{1-\sigma}\right ),& \epsilon > 0,\\
		r^{1+\sigma}, & \epsilon <0,
	\end{cases}
	\label{Eq:tau1slt1}
\end{align}
where $\add$ does not play any role.
Note that Eq.~\eqref{Eq:tau1slt1} reproduces
the exact result for $\sigma=0$ if we set $C_0 =\ln(1+\epsilon)-\ln (1-\epsilon)$.
Although we arrive at Eq.~\eqref{Eq:tau1slt1} by an approximation, 
this result is actually exact when it comes to the leading asymptotic behavior.

\begin{figure*}
	\includegraphics[width=0.9\linewidth]{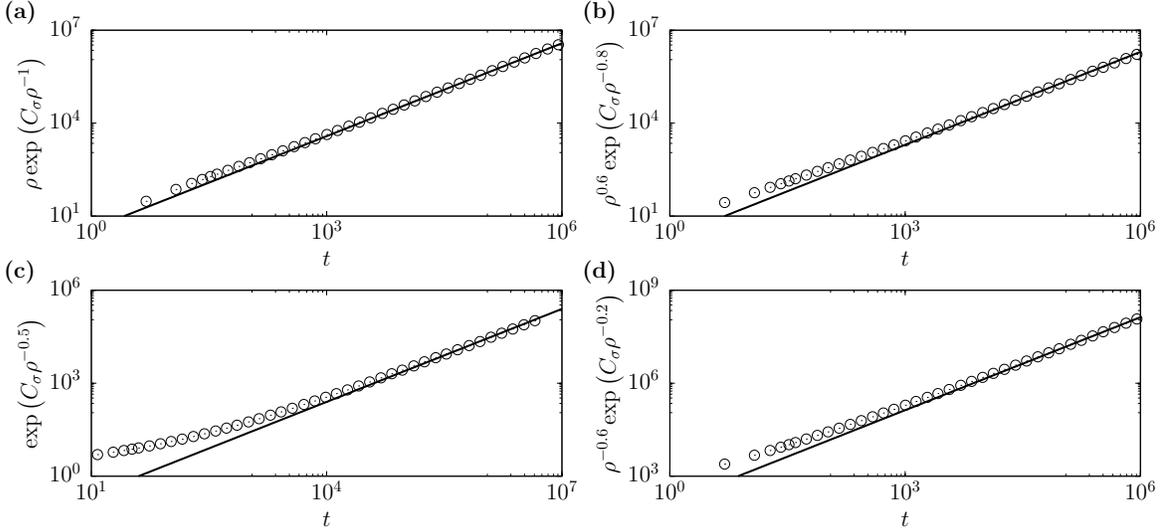}
	\caption{\label{Fig:slt1} Double logarithmic plots of $\rho^{1-2\sigma} \exp \left ( C_\sigma \rho^{-1+\sigma}
	\right )$
	vs. $t$ for (a) $\sigma=0$, $\epsilon = \frac{1}{2}$ ($C_\sigma=\ln 3$),
	(b) $\sigma = 0.2$, $\epsilon = \frac{1}{2}$ ($C_\sigma = 1.25$),
	(c) $\sigma = 0.5$, $\epsilon = \frac{1}{8}$ ($C_\sigma =0.5$),
	(d) $\sigma = 0.8$, $\epsilon = \frac{1}{2}$ ($C_\sigma =5$); see Eq.~\eqref{Eq:delslt1}.
	For guides to the eyes, we also draw a straight line with slope 1 in each panel.
	}
\end{figure*}
Now we investigate the long-time behavior of the \narw  by
analyzing Eq.~\eqref{Eq:drdt} with $\tau(\rho)  = T_0(1/\rho)$.
For $0 \le \sigma < 1$ and $\epsilon>0$, we have
\begin{align}
	\frac{d\rho}{dt} \sim - \rho^{1+\sigma} e^{-C_\sigma\rho^{-1+\sigma}},
\end{align}
which can be solved approximately for small $\rho$ (for large $t$) as
\begin{align}
	\nonumber
	t(\rho) 
	&\sim  \int_\rho^1 \rho^{-1-\sigma} \exp \left ( C_\sigma \rho^{-1+\sigma} \right ) d\rho\\
	\nonumber
	&=  \int_1^{1/\rho} x^{\sigma-1} \exp \left ( C_\sigma x^{1-\sigma} \right ) dx\\
	 &\sim 
	 \rho^{1-2\sigma} \exp \left (C_\sigma\rho^{-1+\sigma}\right ) ,
	 \label{Eq:delslt1}
\end{align}
where we have used Eq.~\eqref{Eq:asym}.  Here
\begin{align}
	C_\sigma = \begin{cases}2\epsilon/(1-\sigma), & \sigma>0,\\
		\ln [(1+\epsilon)/(1-\epsilon)], & \sigma=0.
\end{cases}
\end{align}
Accordingly, we get
\begin{align}
	\nonumber
	\rho(t) &\sim 
	\left [ \ln t - (1-2\sigma) \ln \rho \right ]^{-1/(1-\sigma)}\\
	&\sim (\ln t)^{-1/(1-\sigma)}.
	\label{Eq:rtslt1}
\end{align}

To confirm the prediction, we performed Monte Carlo simulations for $\sigma=0, 0.2, 0.5$ and $0.8$
with system size $L=2^{22}$ ($\sigma=0$) or $L=2^{20}$ ($\sigma\ge 0.2$).
In Fig.~\ref{Fig:slt1}, we compare simulation results
with Eq.~\eqref{Eq:delslt1}.
Our prediction is in excellent agreement with simulations up to nonuniversal multiplication factors. 

\begin{figure*}
	\includegraphics[width=0.9\linewidth]{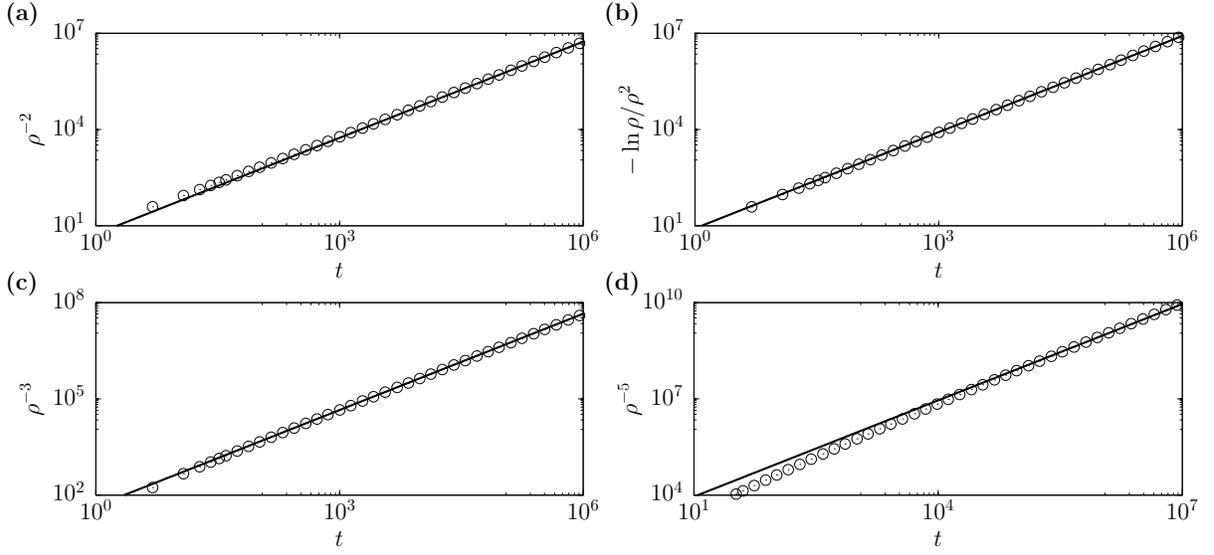}
	\caption{\label{Fig:seq1} Behavior of density of the \narwr with $\sigma=1$ on a double-logarithmic scale.
	(a) $\rho^{-2}$ vs. $t$ for $\epsilon =\frac{1}{4}$.
	(b) $-\ln \rho /\rho^{2}$ vs. $t$ for $\epsilon =\frac{1}{2}$.
	(c) $\rho^{-3}$ vs. $t$ for $\epsilon =1$.
	(c) $\rho^{-5}$ vs. $t$ for $\epsilon =2$.
	For guides to the eyes, we also draw a straight line with slope 1 in each panel.
	}
\end{figure*}
Before closing this subsection, we consider the case with negative $\epsilon$.
Since $\tau(\rho) \sim \rho^{-1-\sigma}$ for $0 \le \sigma < 1$,
we get
\begin{align}
	\frac{d\rho}{dt} \sim -\rho^{2+\sigma}
	\rightarrow \rho \sim t^{-1/(1+\sigma)},
\end{align}
which was already confirmed numerically in Ref.~\cite{Park2020B}; see also Eq.~\eqref{Eq:negdel}.

\subsection{\label{Sec:sigmaeq1}\narw for $\sigma=1$}
Since Eq.~\eqref{Eq:rtslt1} breaks down when $\sigma = 1$, 
we treat the case with $\sigma=1$ separately in this subsection.
Using the approximation~\eqref{Eq:prod_app}, we get
\begin{align}
	\prod_{k=n}^i \frac{1+\epsilon(k+\add)^{-1}}{1-\epsilon(k+\add)^{-1}} 
	\sim 
	\left (\frac{i+\add}{n+\add} \right )^{2\epsilon},
	\label{Eq:s1prodapp}
\end{align}
which gives
\begin{align}
	\nonumber
	T_0 &\sim \int^{r} dx (x+\add)^{2\epsilon}
	\int_1^x (n+\add)^{-2\epsilon} dn\\
	&\sim
	\begin{cases} r^2 , & 2 \epsilon< 1, \\
		r^2 \ln r, & 2 \epsilon =1,\\
		r^{1+2\epsilon}, &2  \epsilon >1.
	\end{cases}
	\label{Eq:T0s1}
\end{align}
As in Sec.~\ref{Sec:Sur}, a logarithmic behavior appears for $2\epsilon = 1$.

Actually, we found exact expressions of $T_0$ for $\sigma=1$.
For $2\epsilon=1$, we find
\begin{align}
	T_0 = r + 2 \sum_{n=1}^{r} \sum_{i=n}^{r-1} 
	\frac{2i+2\add - 1}{2n + 2\add - 1} \sim r^2 \ln r,
\end{align}
and for $2\epsilon \neq 1$
\begin{align}
	\nonumber
	T_0 =
	& \frac{2 \Gamma(\add +1 -\epsilon)}{(4\epsilon^2-1)\Gamma(\add+\epsilon)} 
	\frac{\Gamma(r+\add+1+\epsilon)}{\Gamma(r+\add - \epsilon)}\\
	&+ \frac{r^2 + 2 \add r + 2(\add^2 - \epsilon^2)/(1+2\epsilon)}{1-2\epsilon},
\end{align}
where we have repeatedly used Eq.~\eqref{Eq:diff1}. 
Using the Stirling's formula, one can arrive at Eq.~\eqref{Eq:T0s1}.

Equations~\eqref{Eq:drdt} and \eqref{Eq:T0s1} now yield
\begin{align}
	\frac{d\rho}{dt} \sim 
	\begin{cases} - \rho^3 , & 2 \epsilon<1 \\
		\rho^3/ \ln \rho, & 2 \epsilon = 1,\\
		- \rho^{2+2\epsilon}, & 2 \epsilon > 1,
	\end{cases}
\end{align}
whose solutions are
\begin{align}
	\label{Eq:rhoseq1}
	t \sim \begin{cases}
		\rho^{-2} , & 2 \epsilon<1 \\
		-\ln \rho/\rho^2, & 2 \epsilon = 1,\\
		\rho^{-1-2\epsilon}, &2  \epsilon > 1.
	\end{cases}
\end{align}
Inverting the function, we get the asymptotic behavior of $\rho$ as
\begin{align}
	\rho \sim 
	\begin{cases}
		t^{-1/2}, & 2 \epsilon<1 \\
		\sqrt{\ln t/t}, & 2 \epsilon=1,\\
		t^{-1/(1+2\epsilon)}, & 2 \epsilon > 1.
	\end{cases}
	\label{Eq:rs1}
\end{align}

Now we present our simulation results for four cases with 
$\epsilon = \frac{1}{4}$ ($L=2^{22}$),
$\epsilon = \frac{1}{2}$ ($L=2^{22}$),
$\epsilon = 1$ ($L=2^{22}$), and
$\epsilon = 2$ ($L=2^{21}$).
We use $\add=2$ for $\epsilon\ge 1$ and $\add=0$ for $\epsilon<1$.
The simulation results are presented in Fig.~\ref{Fig:seq1}.
The long-time behavior is in excellent agreement with
our prediction up to nonuniversal multiplication constants.

\subsection{\narw for $\sigma>1$}
Since 
\begin{align}
	\ln \frac{1+\epsilon x}{1-\epsilon x} \le x \ln \frac{1+\epsilon}{1-\epsilon} 
\end{align}
for $0<\epsilon <1$ and
\begin{align}
	\ln \frac{1+\epsilon x}{1-\epsilon x} \le 
	2 \epsilon x
	\label{Eq:ineqneg}
\end{align}
for $-1<\epsilon<0$, where $0<x < 1$,
we have an inequality
\begin{align}
	\nonumber
	\prod_k \frac{1+\epsilon (k+\add)^{-\sigma}}{1-\epsilon (k+\add)^{-\sigma}}
	\le \exp \left [ D_\epsilon \sum_{k=1}^\infty (k+\add)^{-\sigma}\right ] \\
	\le \exp \left [ D_\epsilon \sum_{k=1}^\infty k^{-\sigma}\right ] 
	= \exp \left [ D_\epsilon \zeta(\sigma)\right ],
	\label{Eq:Depsilon}
\end{align}
where $\zeta(\sigma)$ is the Riemann zeta function ($\sigma>1$) and 
\begin{align}
	D_\epsilon = \begin{cases}
		\ln [(1+\epsilon)/(1-\epsilon)],& \epsilon > 0, \\
		2\epsilon, & \epsilon<0.
	\end{cases}
\end{align}
Thus, $T_0$ is bounded by a square function of  $r$.

Since $T_0$ for given $r$ is an increasing (a decreasing) function of 
$\sigma$ for negative (positive) $\epsilon$, we have a lower bound
\begin{align}
	T_0(\sigma>1) \ge \begin{cases}
		T_0(\sigma=\infty)\sim r^2, &\epsilon >0,\\
		T_0(\sigma=1)\sim r^2,& \epsilon<0.
	\end{cases}
\end{align}
Therefore, we get $T_0 \sim r^2$ for any $\epsilon$ if $\sigma>1$
and the \narw with $\sigma >1$ shares the (universal) long-time behavior
with the annihilating random walk without bias.
The same conclusion was arrived at in Sec.~\ref{Sec:Sur}.

\begin{table*}
	\caption{\label{Table:I} The asymptotic behaviors of $R(t)$, $S(t)$, and $\rho(t)$ for $\sigma\le 1$.}
\begin{ruledtabular}
	\begin{tabular}{lcccccc}
		       & \multicolumn{3}{c}{$\sigma=1$} & \multicolumn{1}{c}{$\epsilon>0$} & \multicolumn{2}{c}{$\epsilon<0$}\\
		       \cline{2-4}\cline{5-5}\cline{6-7}
		       & $2\epsilon>1$ & $2\epsilon=1$ & $2\epsilon<1$ & $0\le \sigma < 1$ & $\sigma=0$ & $0<\sigma<1$ \footnotemark[1]\\
		       \hline
		$R(t)$ & $\sqrt{t}$    & $\sqrt{t}$    & $\sqrt{t}$ & $t^{1/(1+\sigma)}$ & constant &  $t^{\gamma} $\\
		$S(t)$&  constant & $(\ln t)^{-1}$ & $t^{-(1-2\epsilon)}$ & constant & $t^{-3/2} \exp[-(1-\sqrt{1-\epsilon^2})t]$ & $t^{-\alpha} \exp[-\lambda t^{\beta} ]$ \\
		$\rho(t)$& $t^{-1/(1+2\epsilon)}$ & $\sqrt{\ln t/t}$ & $t^{-1/2}$ &$(\ln t)^{-1/(1-\sigma)} $ & $t^{-1}$& $t^{-1/(1+\sigma)}$\\
	\end{tabular}
\end{ruledtabular}
	\footnotetext[1]{$\gamma>0$ and $0<\beta<1$. Exact formulas for 
	$\alpha$, $\beta$, $\gamma$, and $\lambda$ are not available in this work.}
\end{table*}
\section{\label{Sec:sum}Summary and discussion}
We have studied the annihilating random walk with long-range interaction in one dimension.
The long-range interaction manifests its presence by the hopping bias 
in the transition rate~\eqref{Eq:qdef}.
We have investigated the survival probability $S(t)$ and the mean spreading $R(t)$
of surviving samples for the two-particle initial condition,
and the density $\rho$ for the fully occupied initial condition.
The results for $\sigma \le 1$ are summarized in Table~\ref{Table:I}.

For $\sigma>1$, 
the system turned out to show the same universal behavior as the unbiased annihilating random walk,
which was already anticipated in Ref.~\cite{Park2020B} for the \narwa.

For $\sigma<1$, the sign of $\epsilon$ plays an important role.
When $\epsilon>0$ (\narwr), we have found
that $S(t)$ saturates to a nonzero value, 
a mean-field-like approximation gives the exact asymptotic behavior of $R(t)$,
and $\rho(t)$ decays logarithmically.
When $\epsilon<0$ (\narwa), the mean-field-like
approximation failed to predict the right asymptotic behavior
for $R(t)$ and $S(t)$ for $0<\sigma<1$. We only reported the numerical results.
But, when $\sigma=0$ and $\epsilon<0$, the exact asymptotic behaviors of $R(t)$ and $S(t)$
are available. Actually, there is a quasistationary state in this case.

For $\sigma=1$, the threshold value of $\epsilon$ is shifted
to $\frac{1}{2}$. When $2 \epsilon > 1$, $S(t)$ saturates to
a nonzero value, while $\rho(t)$ decays with continuously
varying exponent that depends on $\epsilon$.
When $2 \epsilon<1$, $S(t)$ decays with continuously varying
exponent with $\epsilon$, while $\rho(t)$ shows a universal behavior.
When $2\epsilon = 1$, $S(t)$ decays logarithmically and
$\rho(t)$ has a logarithmic correction.
In all cases, $R(t)$ shows the universal $\sqrt{t}$ behavior.

In a different context, continuously varying decaying exponent in coarsening dynamics
was observed in Refs.~\cite{Kim2013,Kim2015}.
We hope our results shed some light on deeper understanding
of the coarsening dynamics in Refs.~\cite{Kim2013,Kim2015}.

For the unbiased case, the annihilating random walk was analyzed by the RG~\cite{Peliti1986,Lee1994}.
It would be an intriguing task to analyze the \narw by the RG,
because the long-range interaction would appear as a multiplication of 
many fields in field-theoretical action.

When branching is introduced to the \narwa, rich critical phenomena 
have been reported especially for the case of the even number of 
offspring~\cite{DR2019,Park2020B,Park2020A}.
In this context, it is natural to ask what would happen if branching
is introduced to the \narwr.
If we think naively, then we would conclude that  
as soon as branching is introduced, 
the steady-state density is nonzero for $\sigma<1$ and $\epsilon>0$, because
$P_s$ is nonzero for $\sigma<1$ in the \narwr. 
Our preliminary studies show, however, that this scenario is not true in general
and the branching actually triggers rich phenomena.
These results will be published elsewhere.

\begin{acknowledgments}
This work was supported by the National Research Foundation of Korea (NRF) grant funded by the Korea government (MSIT) (Grant No. 2020R1F1A1077065) and by the Catholic University of Korea, research fund 2020. 
The author furthermore thanks the Regional Computing Center of the
University of Cologne (RRZK) for providing computing time on the DFG-funded High
Performance Computing (HPC) system CHEOPS.
\end{acknowledgments}
\appendix
\section{\label{App:Gn} Convergence or divergence of $G_n$}
In this Appendix, we prove that
$G_n$ defined in Eq.~\eqref{Eq:F1rform} converges as $n\rightarrow \infty$ if $\sigma<1$ and
$\epsilon>0$ and diverges if $\sigma>1$.

We first consider the case with $\sigma<1$ and $\epsilon >0$.
Using the inequality ($0\le y < 1$)
\begin{align}
	\ln \frac{1-y}{1+y} 
	\le -2 y,
\end{align}
we get
\begin{align}
\prod_{k=1}^i \frac{1-\epsilon (k+\add)^{-\sigma}}{1+\epsilon (k+\add)^{-\sigma}}
	\le \exp\left ( - 2\epsilon\sum_{k=1}^i (k+\add)^{-\sigma}
	\right ).
\end{align}
Since
\begin{align}
	\int_1^i (k+\add)^{-\sigma} dk \le \sum_{k=1}^i (k+\add)^{-\sigma},
\end{align}
we have an inequality
\begin{align}
	\prod_{k=1}^i
	\frac{1-\epsilon(k+\add)^{-\sigma}}{1+\epsilon(k+\add)^{-\sigma}}
	\le C_1 \exp \left [ - C_0(i+\add)^{1-\sigma}
	 \right],
	 \label{Eq:C0C1}
\end{align}
where $C_0 = 2\epsilon/(1-\sigma)$ and 
$C_1 = \exp \left [ (1+\add)^{1-\sigma} C_0 \right ]$.
Since the sum of the right-hand side of Eq.~\eqref{Eq:C0C1} from $i=1$ to $i=\infty$
is obviously finite, 
$G_n$ for positive $\epsilon$ should converge to a finite value as $n\rightarrow \infty$.

Now we move on to the case with $\sigma>1$. Since
\begin{align}
	\ln \frac{1-y}{1+y} \ge -2 y
\end{align}
for $-1<y<0$ and
\begin{align}
	\ln \frac{1-\epsilon y}{1+\epsilon y} \ge  - y \ln \frac{1+\epsilon}{1-\epsilon},
\end{align}
for $0<y<1$ and $0<\epsilon<1$,
there is a positive $C_2$ such that
\begin{align}
	\ln \frac{1-\epsilon(k+\mu)^{-\sigma}}{1+\epsilon(k+\mu)^{-\sigma}}
	\ge -C_2 (k+\mu)^{-\sigma},
\end{align}
for given $\epsilon$.
Since 
\begin{align}
	(1+\add)^{-\sigma} + \int_1^i (k+\add)^{-\sigma} dk \ge \sum_{k=1}^i (k+\add)^{-\sigma},
\end{align}
there are positive constants $C_3$ and $C_4$ such that
\begin{align}
	\prod_{k=1}^i
	\frac{1-\epsilon(k+\add)^{-\sigma}}{1+\epsilon(k+\add)^{-\sigma}}
	\ge C_3 \exp \left [ - C_4(i+\add)^{1-\sigma}
	 \right].
	 \label{Eq:C3C4}
\end{align}
If $\sigma>1$, then the lower bound of Eq.~\eqref{Eq:C3C4} can be set
$C_5= C_3 \exp[-C_4 (1+\add)^{1-\sigma} ]$, which gives $G_n \ge C_5 (n-1)$.
Thus, $G_n$ diverges for any $\epsilon$ if $\sigma>1$.
In a similar manner, one can easily show that there is a positive $C_6$ such that
$G_n \le C_6 n$. Hence, $G_n \sim n$ for $\sigma>1$.

\section{\label{App:asym} Asymptotic expansion using integration by parts}
In this Appendix, we find the leading behavior for large $r$ of the integral
(for a general discussion, see, for example, Ref.~\cite{B:Bruijn})
\begin{align}
	I_1 \equiv A \alpha \int_1^r x^\beta \exp\left (A x^{\alpha}\right ) dx
	=  A\int_1^{r^{\alpha}} y^{\gamma} e^{A y} dy,
\end{align}
where $\alpha > 0$, $A>0$, and $\gamma = (1+\beta-\alpha)/\alpha$.
By an integration by parts , we get
\begin{align}
	I_1 
	&=  r^{1+\beta-\alpha} \exp \left (A r^\alpha\right ) - e^{A} - 
	 \gamma I_2,\\
	I_2 &\equiv \int_{1}^{r^\alpha} y^{\gamma-1} e^{Ay} dy.
\end{align}

For $I_2$, we split the integral as
\begin{align}
	\nonumber
	&\int_1^{r^\alpha/2} y^{\gamma-1} e^{Ay} dy
	\le \frac{1}{A} \max\left \{1,\left (\frac{r^\alpha}{2}\right )^{\gamma-1}\right \} e^{Ar^\alpha/2},\\
	&\int_{r^\alpha/2}^{r^\alpha} y^{\gamma-1} e^{Ay} dy
	\le \frac{1}{A} \max\{1,2^{1-\gamma}\} r^{1+\beta-2\alpha} e^{Ar^\alpha},
\end{align}
which shows $I_2/I_1 \rightarrow 0$ as $r\rightarrow 0$.
Hence, we get
\begin{align}
	I_1 \sim r^{1+\beta-\alpha} \exp \left (A r^\alpha\right ).
	\label{Eq:asym}
\end{align}

\section{\label{App:s0att}Derivation of Eq.~\eqref{Eq:Pits0}}
In this Appendix, we derive Eq.~\eqref{Eq:Pits0} for $\sigma=0$.
We first find $d_{i,n}$ defined in Eq.~\eqref{Eq:dindef}.
Let $q$ be the probability of hopping to the right.
For the walker to arrive at site $i$ after $n$ jumps, the number of hopping to the right 
should be $(n+i-1)/2$, where $n+i$ must be an odd number and $1\le i \le n+1$.
Since the probability of hopping does not depend on site index $i$, 
we can write
\begin{align}
	d_{i,n} = M_{i,n} q^{(n+i-1)/2} (1-q)^{(n-i+1)/2},
\label{Eq:din}
\end{align}
where $M_{i,n}$ is the number of paths that do not meet the absorbing wall. 
Using the reflection principle of random-walk paths~\cite[p. 72]{FellerI}, we get
\begin{align}
M_{i,n} &= \binom{n}{(n+i-1)/2} - \binom{n}{(n+i+1)/2}\nonumber \\
&= \frac{n! i}{[(n+i+1)/2]! [ (n-i+1)/2]!}.
\label{Eq:Min}
\end{align}

Plugging Eq.~\eqref{Eq:din} with Eq.~\eqref{Eq:Min} into Eq.~\eqref{Eq:Pin}, 
we get for $i=2k-1$ ($n=2m$)
\begin{align}
\nonumber
& P_{2k-1}(t) 
	= ie^{-t} \sum_{m=k-1}^\infty \frac{q^{m+k-1} (1-q)^{m-k+1} t^{2m}}{(m+k)! (m-k+1)!}\\
	&=\frac{iw^{i}}{qt} e^{-t}
	\sum_{m=0}^\infty \frac{(x/2)^{2m+2k-1}}{(m+2k-1)! m!}= \frac{iw^{i}}{qt} e^{-t}I_{i} (x),
\end{align}
and for $i=2k$ ($n=2m+1$) 
\begin{align}
\nonumber
&P_{2k}(t) 
= i e^{-t} \sum_{m=k-1}^\infty \frac{q^{m+k} (1-q)^{m-k+1} t^{2m+1}}{(m+k+1)! (m-k+1)!}\\
	&=\frac{iw^{i}}{qt}e^{-t}	\sum_{m=0}^\infty \frac{(x/2)^{2m+2k}}{(m+2k)! m!} 
	= \frac{iw^{i}}{qt} e^{-t} I_{i} (x),
\end{align}
where $w = \sqrt{q/(1-q)}$ and $x=2 t \sqrt{q(1-q)}$.
Thus, 
\begin{align}
	P_i(t) = \left ( \sqrt{\frac{q}{1-q}} \right )^{i-1} \frac{2i}{x}I_i (x) e^{-t}
\end{align}
is valid for all $i \ge 1$.
Putting $q=(1+\epsilon)/2$, we get Eq.~\eqref{Eq:Pits0}.
\bibliography{Park}
\end{document}